\documentclass[
]{ceurart}

\usepackage{listings}
\usepackage{todonotes}
\usepackage{xcolor}
\usepackage[nolist]{acronym}
\usepackage{verbatim}
\usepackage{xspace}
\usepackage{multirow}
\usepackage{graphicx}

\newcommand{\ie}{\textit{i}.\textit{e}.,\xspace}

\begin{document}

\copyrightyear{2023}
\copyrightclause{Copyright for this paper by its authors.
  Use permitted under Creative Commons License Attribution 4.0
  International (CC BY 4.0).}

\conference{Italian Workshop on Artificial Intelligence for Human Machine Interaction (AIxHMI 2023), November 06, 2023, Rome, Italy}

\title{An experimental protocol to access immersiveness in video games}



\author[1]{Marika Malaspina}[%
orcid=,
email=m.malaspina11@campus.unimib.it
]

\author[1]{Jessica Amianto Barbato}[%
orcid=0000-0002-9831-3210,
email=jessica.amiantobarbato@unimib.it
]

\author[1]{Marco Cremaschi}[%
orcid=0000-0001-7840-6228,
email=marco.cremaschi@unimib.it,
url=,
]
\cormark[1]

\author[1,2]{Francesca Gasparini}[%
orcid=0000-0002-6279-6660,
email=francesca.gasparini@unimib.it,
url=https://mmsp.unimib.it/francesca-gasparini/,
]

\author[1,2]{Alessandra Grossi}[%
orcid=0000-0003-1308-8497,
email=alessandra.grossi@unimib.it,
url=https://mmsp.unimib.it/alessandra-grossi/,
]

\author[1,2]{Aurora Saibene}[%
orcid=0000-0002-4405-8234,
email=aurora.saibene@unimib.it,
url=https://mmsp.unimib.it/aurora-saibene/,
]

\address[1]{University of Milano-Bicocca, Viale Sarca 336, 20126, Milano, Italy}
\address[2]{NeuroMI, Milan Center for Neuroscience, Piazza dell’Ateneo Nuovo 1, 20126, Milano, Italy}

\cortext[1]{Corresponding author.}

\begin{abstract}
  In the video game industry, great importance is given to the experience that the user has while playing a game. In particular, this experience benefits from the players' perceived sense of being in the game or \textit{immersion}. 
  The level of user immersion depends not only on the game's content but also on how the game is displayed, thus on its \ac{ui} and the \ac{hud}.
  Another factor influencing immersiveness that has been found in the literature is the player's expertise: the more experience the user has with a specific game, the less they need information on the screen to be immersed in the game.\\
  Player's level of immersion can be accessed by using both questionnaires of their perceived experience and exploiting their behavioural and physiological responses while playing the target game.\\
  Therefore, in this paper, we propose an experimental protocol to access immersiveness of gamers while playing a third-person shooter (Fortnite) with UIs with a standard, a dietetic, and a proposed HUD. 
  A subjective evaluation of the immersion will be provided by completing the \ac{ieq}, while objective indicators will be provided by face tracking, behaviour and physiological responses analyses.\\
  The ultimate goal of this study is to define guidelines for video game UI development that can enhance the players' immersion.
\end{abstract}


\maketitle


\section{Introduction}\label{sec:intro}

In the gaming sector, \textbf{immersion} is an element considered an important interaction experience \cite{Brown2004}. In literature, the concept of immersion is generally reconnected and defined through the concept of \textbf{flow}, settled by the psychologist \textit{Mihály Csikszentmihalyi} in 1975 \cite{csikszentmihalyi1975beyond, Csikszentmihalyi1990}. The author has described flow as an \emph{optimal experience}, thus an experience which is achieved by people when they enter into a state of euphoria, characterised by total attention on what they are doing, by loss of sense of self, and by an alteration of time perception. 

While \textit{Csikszentmihalyi}'s studies were performed on different types of people, such as sportsmen and artists, lately, the concept has been translated to other fields. In particular, considering video games, the literature presents different opinions about the veracity overlap of the flow and immersion concepts \cite{Brown2004,Jennett2008}, where immersion can be simply described in this case as ``\emph{the sense of being in the game}''. This means that immersion can be considered not only as a feeling of presence in the game but also as a real cognitive statement whereby people are immersed in their activity (but not as much as in flow) \cite{Taylor2017}. 

The subject of immersion in the video game context has prompted numerous studies and projects. For example, in \textit{Sanders and Cairns}'s work \cite{sanders&cairns}, video game immersion is analysed considering the timing of the game session and the use of music, while \textit{Christou} \cite{christou2014interplay} provides an analysis of the relationship between immersion and appeal.

A significant aspect of this topic involves analysing this phenomenon starting from the video game \ac{ui}. The \ac{ui} plays a crucial role as it is the point of interaction between the player and the video game. It also acts as a filter for the player's game experience. 

Among the works that focus on the study of immersion in video games by considering their \ac{ui}s, \emph{Iacovides et al.}'s one \cite{Iacovides2015} compares the diegetic and non-diegetic \ac{ui} versions of the case study, which is \emph{Battlefield 3}\footnote{Electronic Arts Inc. (\url{https://www.ea.com/en-gb/games/battlefield/battlefield-3?setLocale=en-gb} accessed September 21, 2023)}, and evaluates if both \ac{ui}s lead to a positive game experience using questionnaires. Afterwards, they examine the involvement experienced by novices and experts for each of the proposed \ac{ui} versions. Notice that, in this case, for \emph{expert} is intended for a player using a specific game or a game category for at least half an hour per week. The authors find that players with expertise can be more immersed when the \ac{hud} is dietetic, \ie when no information is displayed.

In gaming, another aspect that can be investigated to analyse immersivity is the players' physiological response during the gameplay experience. As part of the research on this topic, \textit{Lages} \cite{lages2021opportunities} introduces the concept of \emph{immersive entertainment} to identify all the entertainment applications devised for immersive systems. The author presents eight factors influencing immersive entertainment: (i) presence and user experience understanding, (ii) characters and storytelling, (iii) interaction technique design, (iv) virtual environment rendering, (v) physiological sensing and biofeedback introduction, (vi) social experience improvement, (vii) user safeguarding and responsible design, and (viii) content creation tool availability.

In particular, \textit{Lages} points out for factor (v), on which we are particularly interested, that there is currently (2021, year of publication) a vast use of facial feature trackers, but that newer headsets integrating these sensors with physiological ones, \ie \ac{eeg}, \ac{eog}, \ac{emg}, \ac{eda}, and \ac{ppg}, are being developed, especially to detect users' emotional states.

On this topic, \textit{Hughes and Jorda} \cite{hughes2021applications} present biosignals commonly used in research on gaming, especially for rehabilitation purposes and/or targeting people with disabilities. In particular, the authors find that mostly EMG, EEG, EDA, eye tracking with EOG, and \ac{ecg} non-invasive technologies are employed considering both user monitoring (passive mode) and affective gaming (active mode).
Considering the provided analyses, EMG is involved in an active modality, and thus for game control or rehabilitation exercises; EEG is efficient for emotion detection and neurological assessment in a passive mode, while its signals can be used actively for game control and virtual environment change; eye tracking and EOG provide information on the user's fatigue, the points of interest in the visualised content, and more actively these signals can be used to control games; ECG allows stress monitoring and the level of engagement, as well as changes of environment and gameplay experience according to the monitored signals; these changes can be also enabled by analysing the EDA signals, which are good indicators of stress, panic, and other negative emotions.

Despite the extensive research conducted using physiological signals, their implementation within the context of video gaming remains limited. 
\textit{Bian et al.} \cite{bian2016framework} propose a model based on the idea that flow is a state of moderate arousal accompanied by a state of joy. For this reason, physiological parameters related to these processes can predict the flow state. In the proposed model, five indicators are included, \ie facial EMG, cardiovascular activity, EDA, respiratory activity, and EEG, accompanied by their features. Notice that this model is also based on a previous study \cite{tozman2015understanding}, which found the connection between flow and heart rate variability using a driving simulator and is applied in the context of \ac{vr}.

Therefore, inspired by \textit{Iacovides et al.} and \textit{Bian et al.} works, we propose an experimental protocol to evaluate immersion, considering the characteristics of \ac{ui}s in the \emph{Fortnite}\footnote{Epic Games Inc. (\url{https://store.epicgames.com/en-US/p/fortnite?lang=en-US} accessed September 19, 2023).} video game, on the level of expertise of Fortnite users as well as on their physiological responses.  

In particular, the \ac{ieq} introduced by \textit{Iacovides et al.} will be used to evaluate the perceived subjects' level of immersion. Moreover, face tracking, EEG, \ac{gsr}, PPG, and respiratory data will be collected for a physiological evaluation of the phenomenon under analysis, following guidelines made from \textit{Bian et al.}'s physiological model. 

This study aims to define guidelines for creating video game \ac{ui}s that can optimise the immersion of expert and novice players within the target video game category, \ie \ac{tps}. 
Thus, the research coming from the proposed experimental protocol is intended to answer the following research questions:
\begin{itemize}
    \item How can we discriminate different levels of immersiveness? 
    \item Do the physiological signals contribute to the interpretation of the levels of immersiveness considering them separately or as an ensemble? 
    \item Does the players' expertise influence the outcome, and in what measure? 
    \item Does the information present on the UI affect the players' immersiveness? 
\end{itemize}

The rest of the paper is organised as follows. Section \ref{sec:role-of-ai} provides a brief overview of the role of \ac{ai} in physiological signal interpretation, especially in the fields of affective computing and gaming. Section \ref{sec:experimental-settings} and Section \ref{sec:experimental-protocol} report the experimental settings and protocols. Finally, conclusions are drawn in Section \ref{sec:conclusion}.

\section{The role of artificial intelligence}
\label{sec:role-of-ai}
Video games elicit emotions in their players, influencing their sense of being in the game \cite{setiono2021enhancing}. Moreover, the player's behaviour directly influences the gameplay and game objectives. 

\textit{Affective computing} automatically recognises human emotions by studying implicit measures derived from physiological sensors and facial expressions, allowing close interaction between humans and machines \cite{picard1999affective, marin2020emotion}.
In fact, \textit{affective models} can be designed to exploit and translate the data collected from the previously cited sensors (Section \ref{sec:intro}). This enables computer systems to perceive and interpret human emotions and consequently provide intelligent, empathetic, and responsive interactions \cite{Tao2006}.


Therefore, affective computing can be leveraged to dynamically change a video game, adjusting the storyline by identifying a player's emotional state and actions. This particular application, which may also include game control by means of physiological signals, is called \textit{affective gaming} \cite{kotsia2013affective} and has as its main aim the creation of more profound and immersive gaming experiences that can evoke strong emotions.

A notable and recent example of this approach is \textit{Before Your Eyes}\footnote{\url{https://www.beforeyoureyesgame.com/} accessed September 20, 2023.}, a narrative-driven adventure game that takes players on an emotional journey through the life of a character named Benjamin. Players navigate the story by blinking their eyes, whose movements are tracked by a webcam or an eye-tracking technology.

Besides being a powerful storytelling medium, the game becomes a means to evoke genuine emotions. In particular, \textit{Before Your Eyes} tries to elicit empathy and understanding of Benjamin's story in the players. This capacity may represent a good starting point to drive social change and promote a more compassionate society. Moreover, affective gaming can represent a potential therapeutic tool, allowing users to explore and process their emotions.

Another important application of affective gaming can be related to the \ac{dda}, which is a mechanism that pushes the player to change the game's difficulty level, considering their performance. This operation may not be performed manually by a player during the gameplay and could additionally demoralise the player, who may feel like is resorting to this expedient to continue the game. Using information from different sensors, it is possible to tune the game difficulty and maximise one of four emotional responses: challenge, competence, flow, and valence \cite{Moon2022}.


Given these premises, affective computing and gaming will be considered in a first phase to evaluate the players' responses to the different UIs proposed in the following experimental protocol, trying to assess users' levels of immersiveness and their perceived emotions. In a second phase, a model will be devised to dynamically change the game UI according to the real-time perceived immersion by the players, exploiting the outcome of the previous phase.

\section{Experimental settings}
\label{sec:experimental-settings}


This Section provides the information related to the experimental settings of the proposed experimental protocol, which is divided into three main phases (detailed in Section \ref{sec:experimental-protocol}). Therefore, a brief description of each questionnaire, physiological sensors, and gaming devices will be presented during the research.

\subsection{Questionnaires \label{subsec:questionnaires}}

The experimental design includes the implementation of a \emph{preliminary questionnaire}, utilised during Phase 0 (Section \ref{subsec:phase-0}), and a \emph{pre-test} and an \emph{IEQ questionnaire}, employed during the Phases 1 (Section  \ref{subsec:phase-1}) and 2 (Section \ref{subsec:phase-2}). 

The initial survey analyses responses from expert gamers\footnote{Expert gamers are people who play the shooter game category at least half an hour a week.}, focusing on their preferences for shooter games. The questionnaire prompts users to identify the elements that require modification, relocation, or exclusion, with the aim of enhancing the gaming experience.  

\begin{figure}
    \centering
    \includegraphics[width=0.75\linewidth]{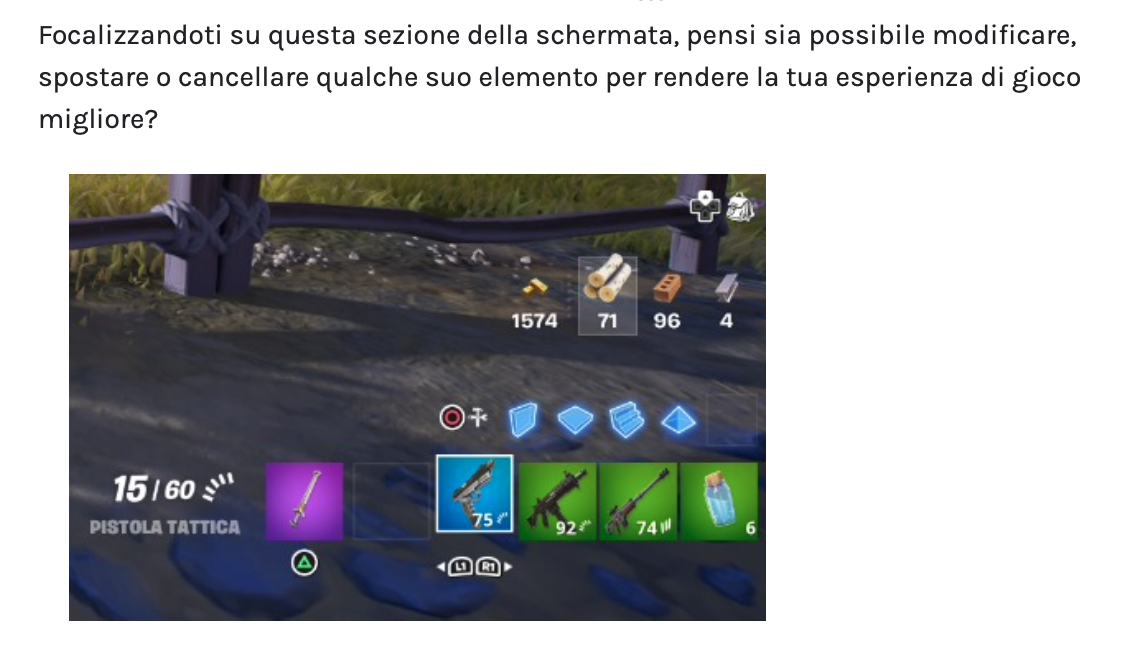}

  \caption{Preliminary questionnaire game frame example with accompanying question: ``Focusing on this game frame, do you think that it is possible to modify, move, or delete some of its elements to improve your gaming experience?''.}
    \label{fig:questionnaire}
\end{figure}

The questionnaire comprises two parts. The first part is a demographic section covering the participant's age, gender, and gaming behaviour. The second part presents various sections of the game frame and the complete game frame of the case study, \ie Fortnite. Participants are requested to comment on potential improvements to individual sections that could enhance their overall gaming experience. Figure \ref{fig:questionnaire} depicts an example of the game frame proposed to the participant and its accompanying question: ``Focusing on this game frame, do you think that it is possible to modify, move, or delete some of its elements to improve your gaming experience?''. Notice that the questionnaire is intended for Italian speakers only, and its purpose is purely to gather feedback from expert users, which will inform adjustments to the \ac{hud} in line with findings from Nielsen's \cite{nielsen} and Federoff's \cite{Federoff2022} heuristics. 

In Phases 1 and 2, other two questionnaires are employed. The first one is a \emph{pre-test} questionnaire, which collects some demographics on the subjects and their gaming expertise. 

The second questionnaire, \ie IEQ, surveys the subject's video game experience during the experimental phases. The IEQ, which was also used in the gaming sector by \emph{Iacovides et al.} \cite{Iacovides2015}, is a questionnaire that reveals the general experience of immersivity \cite{Jennett2008}. IEQ has five dimensions, which are analysed by the different proposed questions. In particular, these dimensions reflect different aspects of an immersive experience, \ie (i) cognitive involvement, (ii) emotional involvement, (iii) real-world dissociation, (iv) challenge, and (v) control. These five interrelated components are evoked by 31 questions, which have a 5 point-scale.

\subsection{Face tracking and behaviour analysis \label{subsec:gopro}} 
A \textit{GoPro HERO10 Black}\footnote{GoPro Inc. (\url{https://gopro.com/en/us/shop/cameras/hero10-black/CHDHX-101-master.html} accessed September 19, 2023).} is used to collect video recordings (5.3K60 + 4K120 video resolution) to be exploited for face tracking and behaviour analysis. In fact, previous research \cite{Wang2016} has demonstrated its ability to provide data on detecting the participant's facial behaviour. This feature permits facial tracking and analyses of the user's visual movements to compensate for the lack of an eye-tracking device \cite{Jennett2008}.

\subsection{Physiological devices \label{subsec:physiological-devices}}
\begin{figure}
    \centering
    \includegraphics[width=0.8\textwidth]{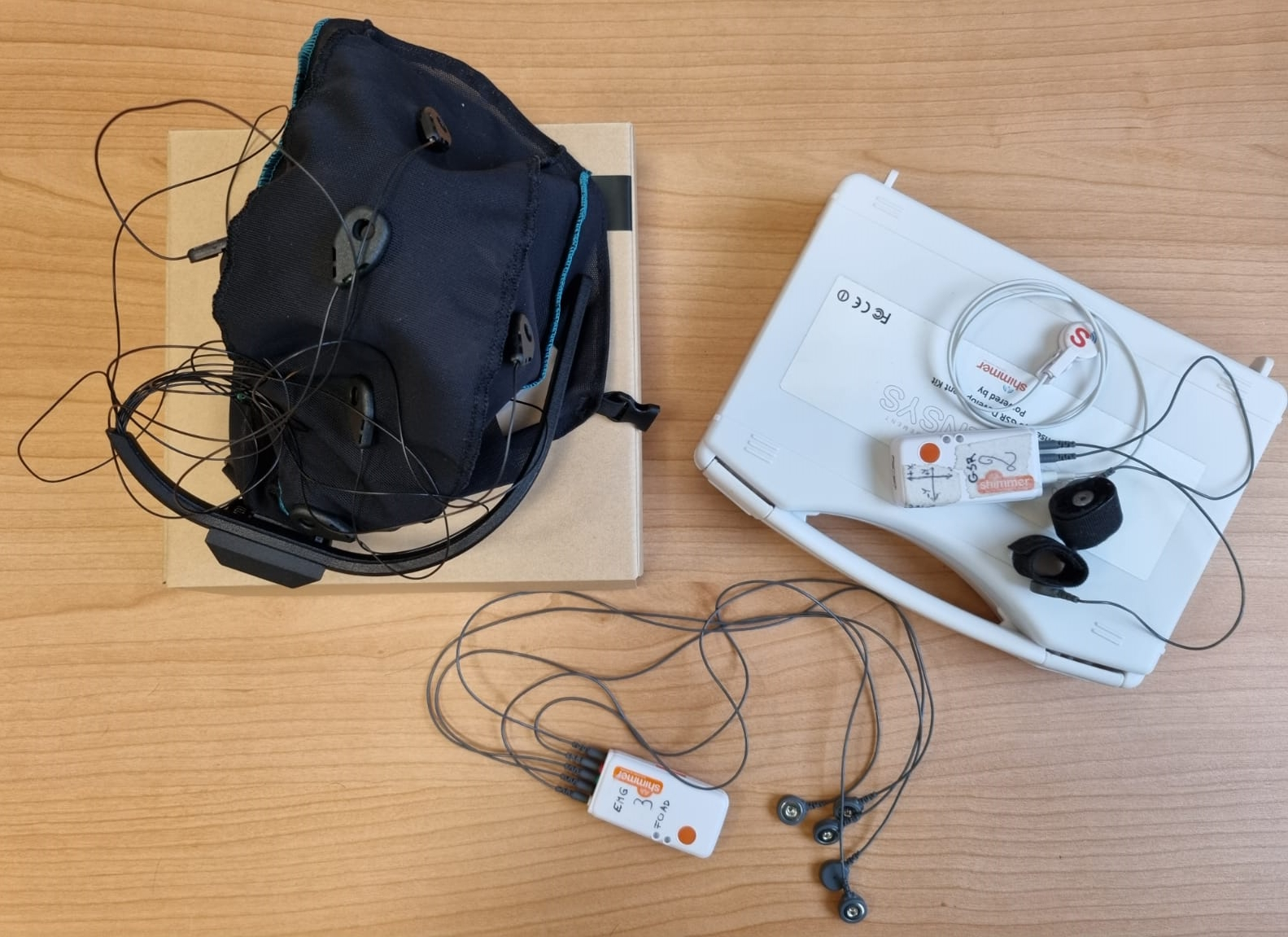}
    \caption{Physiological devices that will be used to record the subjects' physiological responses. From left to right, top to bottom: Unicorn, GSR/PPG device, and respiration device.}
    \label{fig:sensors}
\end{figure}

This Section provides a brief overview of the technical characteristics of the selected physiological devices. Figure \ref{fig:sensors} depicts the EEG (Section \ref{sec:eeg}), the GSR/PPG (Section \ref{sec:gsr}), and the respiration (Section \ref{sec:respiration}) devices.

\subsubsection{Electroencephalographic device}\label{sec:eeg}
The \emph{Unicorn Brain Interface}\footnote{g.tec medical engineering GmbH (\url{https://www.unicorn-bi.com/} accessed August 16, 2023).} (Unicorn) is used to record the EEG signals. This interface comprises the acquisition software, the EEG cap, a default eight electrode set (Fz, C\{3, z, 4\}, Pz, PO\{7, 8\}, and Oz plus ground and reference sensors placed on the subjects' mastoids), and a Bluetooth dongle. This wearable device has the advantage of not relying on wired connections and being characterised by hybrid sensing technologies, \ie the electrodes can be used with a dry or wet configuration.

Notice that all the available electrodes will be employed to capture neural responses elicited by the game events, considering that \textit{Bian et al.} suggest the analyses on the alpha and beta frequency bands only without specifying electrode positioning and that other works \cite{gannouni2021emotion} show that electrodes from different brain areas are selected when emotion recognition is involved. In particular, \textit{Gannouni et al.} \cite{gannouni2021emotion} found that each emotion is coupled with a specific frequency band and a specific electrode set covering the whole brain surface.

Finally, notice that the Unicorn analog-to-digital converter has a resolution of 24 bits and a sampling frequency of 250Hz. 
The Unicorn also presents 3-axis accelerometer and gyroscope. The acceleration and gyroscope ranges are $\pm 8$g and $\pm 1000$°/s setting in x, y, and z directions. The bandwidth is 44.8Hz for the accelerometer while 41Hz for the gyroscope.

\subsubsection{Galvanic skin response and photoplethysmogram}\label{sec:gsr}
GSR and PPG signals are acquired with the \emph{Shimmer3 GSR+ Unit}\footnote{Shimmer (\url{https://shimmersensing.com/product/shimmer3-gsr-unit/} accessed August 16, 2023).} (Shimmer GSR). This device comprises two electrodes to monitor skin conductivity and an optical pulse sensor to be placed on the earlobe. Moreover, it is integrated with 3-axis accelerometer, gyroscope, magnetometer, and altimeter. Both PPG and GSR data are acquired using a sampling frequency of 128Hz.

As for the Unicorn, the Shimmer GSR is a wearable and wireless device based on Bluetooth technology. The device measurement range is 10k-4.7M$\Omega$ (.2uS – 100uS) $\pm$ 10\%. 22k-680k$\Omega$ (1.5-45uS) $\pm$ 3\%, while the frequency range is 15.9Hz.
In the proposed experiment, the device will be placed on the participant's left hand with the two GSR electrodes attached respectively to the index and pinkie fingers to reduce possible artefacts due to the use of the joypad.

\subsubsection{Respiration device}\label{sec:respiration}
The \emph{Shimmer3 EMG Unit}\footnote{Shimmer (\url{https://shimmersensing.com/product/shimmer3-emg-unit/} accessed August 16, 2023).} (Shimmer EMG) records the respiration demodulation through ECG data. The device consists of two pairs of electrodes placed on the midclavicular lines to measure the respiration rate as charging impedance across the chest \cite{redmond2013trans}. In addition, an electrode is placed on the right side of the sternum as a reference. A sampling frequency of 128 Hz is used to acquire the signals.

Notice that Shimmer3 EMG is also a non-invasive device that allows monitoring and collecting data through Bluetooth wireless technology.

\subsection{Gaming device and selected game}
PlayStation\textregistered5\footnote{Sony Interactive Entertainment (\url{https://www.playstation.com/en-gb/ps5/?smcid=pdc\%3Aen-gb\%3Aprimary\%20nav\%3Amsg-hardware\%3Aps5} accessed September 19, 2023).} has been selected as the gaming device with its DualSense\texttrademark wireless controller. 

The selected TPS game is \emph{Fortnite}. The authors of this paper selected this game for two main reasons. Firstly, \emph{Fortnite} is a famous and appreciated game, having more than 350 million registered users by the end of 2022\footnote{\emph{everyeye.it} (\url{https://www.everyeye.it/notizie/giocatori-fortnite-totale-risposta-612779.html} accessed September 21, 2023).}. Secondly, the HUD has a very important role in the game. In fact, \emph{Celia Hodent} and her team devised the HUD to be informative and non-intrusive, considering the whole user experience. According to \emph{Hodent}\footnote{Celia Hodent (\url{https://celiahodent.com/understanding-the-success-of-fortnite-ux/} accessed September 21, 2023).}, information is always available to the players avoiding taxing them with learning and memorising everything, thus avoiding an excessive cognitive load.


\section{Experimental Protocol}
\label{sec:experimental-protocol}

The experimental protocol is divided into three main phases of interface definition, assessment of the immersiveness with expert players, and assessment of usability with non-expert players.

In fact, \textit{Phase 0} is devoted to the definition of a HUD (from now on \textit{proposed HUD}) in-between the standard Fortnite \ac{ui} and the one that does not report any information on the screen to enhance the immersion of the player while avoiding removing useful information to complete the game session.

Before briefly introducing the other phases, notice that the three UIs that will be used in the experiment can be summarised as follows:
\begin{itemize}
    \item The \ac{ui} with Fortnite standard HUD;
    \item The \ac{ui} without any information, thus without the HUD;
    \item The \ac{ui} created using gamers' opinion and the heuristics present in the literature \cite{nielsen, Federoff2022}, called proposed HUD.
\end{itemize}

Notice that the first two interfaces are considered following the example of \emph{Iacovides et al.} \cite{Iacovides2015}. In fact, \emph{Iacovides et al.} compare these two types of interfaces and find that expert users were more immersed when using the \ac{ui} without a HUD, and thus without information that would distract them. This observation did not apply to non-expert players, who did not obtain any increase in immersion probably due to their inexperience in the target game, \ie \emph{Battlefield 3}.

Proceeding with the subsequent phases, \textit{Phase 1} is intended for expert Fortnite players, \ie Fortnite users that have played the game for at least two years and one hour per week. 
This phase will assess the effectiveness of the proposed HUD resulting from Phase 0, comparing it to the case where the participants play the game using the non-diegetic Fortnite HUD.
Instead, \textit{Phase 2} will evaluate the usability of the proposed HUD by non-expert Fortnite players, \ie participants who know how to use a PlayStation console but have never played Fortnite. This is to understand how expertise influences the definition of the HUD and how the HUD adapts to different users. Moreover, immersiveness will also be accessed for non-expert players.

Physiological responses will be collected to evaluate the best interface type for immersiveness optimisation considering the users' experience while playing the game.

Table \ref{tab:phases} summarises the characteristics of each phase and provides information on the participants' inclusion criteria, the questionnaires that will be provided, and the physiological data that will be recorded.

As a final remark, notice that players are left free to choose which hero to play during the experiment, as outfits do not guarantee competitive advantages within the game \cite{schober2020fortnite}.

\begin{table}[h]
  \centering
  \caption{Summary of the three phases of the proposed experimental protocol.}
    \begin{tabular}{lp{80pt}p{80pt}p{80pt}p{80pt}}
    \toprule
    Phase & Description & Inclusion criteria & Methods & Indicators \\
    \midrule
    0     & Interface definition & Participant plays a TPS for at least 30 min per week & Preliminary questionnaire plus Nielsen and Federoff's heuristics & none \\
    1     & Immersiveness assessment with expert participants & Fortnite expert player aged 18-35 y.o., with a mimum of 2 years experience in Fortnite and playing to it at least  1 hour per week & pre-test questionnaire and IEQ & EEG, GSR, PPG, respiration, facial expression \\
    2     & Usability and immersiveness assessment with non-expert participants & Fortnite non-expert player aged 18-35 y.o., using a PlayStation console with a joypad & pre-test questionnaire and IEQ & EEG, GSR, PPG, respiration, facial expression \\
    \bottomrule
    \end{tabular}%
  \label{tab:phases}%
\end{table}%

\subsection{Phase 0: interface definition \label{subsec:phase-0}}

The aim of Phase 0 is to formulate a revised \emph{Fornite} interface with HUD based on the responses obtained from the \textit{preliminary questionnaire} detailed in Section \ref{subsec:questionnaires}. The modifications will thus respond to the preferences of experienced users and the requirements imposed by Nielsen \cite{nielsen} and Federoff's \cite{Federoff2022} heuristics. The proposed HUD will later be presented to expert and non-expert players in Phase 1 and 2, respectively. 

\subsubsection{Inclusion and exclusion criteria}
Only expert players will be involved in this definition phase. In this phase, for \emph{expert} is intended for a user who plays shooter games for at least 30 minutes per week. Notice that in this phase, the user is not necessarily an expert in Fortnite.

\subsubsection{Materials and methods}
The materials and methods used in this phase are:
\begin{itemize}
    \item The 10 usability heuristics of Nielsen \cite{nielsen};
    \item The heuristics for the gaming sector identified and analysed by Federoff \cite{Federoff2022};
    \item A preliminary questionnaire for the expert gamers\footnote{Available at \url{https://docs.google.com/forms/d/e/1FAIpQLSfvCgCW9XxWfKekU5WOlxmAUwVEPqt-oPBktEdfolcWVNdG4g/viewform?usp=sf_link}}.
\end{itemize}

\emph{Unreal Engine}\textregistered\footnote{Epic Games Inc. (\url{https://www.unrealengine.com/en-US} accessed September 21, 2023).} will be used to create the proposed HUD. Notice that this tool is the same developed and used by \emph{Epic Games}, \emph{Fortnite} developer.

\subsection{Phase 1: immersiveness with expert players \label{subsec:phase-1}}

The main aim of Phase 1 is to evaluate the immersion of the proposed HUD obtained in Phase 0 with respect to the case study dietetic HUD version. 

In fact, expert \emph{Fortnite} players will have to play a game session with the UI (i) without HUD and (ii) with the proposed HUD.

The experiment is conducted in a within-subject condition through two gaming sessions carried out randomly by expert players.

So, it is a (2x1) experiment with \emph{p}=0.8 (the same as \emph{Iacovides et al.} \cite{Iacovides2015}), and the number of expert players needed is 34.

\subsubsection{Inclusion and exclusion criteria}
Expert users aged between 18 and 35 who have played Fortnite for at least two years and play it an hour per week will be involved in Phase 1. Expert users who have played shooter games but not Fortnite will be excluded.

\subsubsection{Materials and methods}
In Phase 1, a pre-test questionnaire\footnote{Available at \url{https://docs.google.com/forms/d/e/1FAIpQLSeKRO65yJVSSzps5XrWL9PXB2YDJqg-9ZmuXR7UKzpc3P_4ng/viewform?usp=sf_link}} and an IEQ\footnote{provided by \emph{Jennet et al.} \cite{Jennett2008} at \url{https://docs.google.com/forms/d/e/1FAIpQLSeRMgdk4_IMxZ-LFcrj9le8bjRaclUHq1G80Y1t_OWoHKGmmw/viewform?usp=sf_link}.} after each game sessions will be presented to the participants.

As for \emph{Iacovides et al.} \cite{Iacovides2015}, participant expertise and the number of interfaces (two) will be considered to understand the immersion level of the expert during the game sessions.

Face tracking and player behaviour will be derived from the videos recorded by means of \emph{GoPro HERO10 Black}.
Moreover, the following devices will be used to exploit different physiological indicators, as suggested by  \emph{Bian et al.}'s framework \cite{bian2016framework}: Unicorn, Shimmer GSR, Shimmer EMG. Section \ref{subsec:physiological-devices} reports more details on the devices.

\subsection{Phase 2: usability with non-expert players\label{subsec:phase-2}}

The main aim of Phase 2 is to evaluate the usability of the proposed HUD based on the feedback of expert shooter gamers obtained in Phase 0 with respect to the classic interface of the case study.

In fact, non-expert Fortnite players will have to play a game session with the UI (i) with the standard HUD, and (ii) with the proposed HUD.

The experiment is conducted in a within-subject condition through two random gaming sessions by non-expert players. 

So, it is a (2x1) experiment with \emph{p}=0.8 (the same as \emph{Iacovides et al.} \cite{Iacovides2015}), and the number of non-expert players needed is 34.

Besides accessing the proposed HUD usability, the player level of immersion will also be evaluated.

\subsubsection{Inclusion and exclusion criteria}
Non-expert users who have not played Fortnite aged between 18 and 35 will participate in this experiment phase. These non-expert users are selected among players who know how to play with a PlayStation console. 

\subsubsection{Materials and methods}
As for Phase 1, the pre-test questionnaire and IEQ will be presented to the participants. Moreover, participant expertise and the number of interfaces (two) will be considered to understand the usability of the proposed HUD as well as the immersion level of the non-expert player during the game sessions.

The same devices reported for Phase 1 will be used: \emph{GoPro HERO10 Black} for face tracking and behaviour analysis, and Unicorn, Shimmer GSR, and Shimmer EMG for physiological responses recording.

\section{Conclusions}
\label{sec:conclusion}
In this paper, a brief overview of the experimental protocol intended to assess the immersiveness of expert and non-expert participants while playing a third-person shooter is presented.

The experiment will be carried out by following literature works \cite{Iacovides2015, bian2016framework} that successfully assessed players' immersion levels considering different HUDs and physiological responses.
Moreover, a HUD for the Fortnite UI will be proposed based on expert shooter players' opinions to enhance the gamers' immersiveness.

Affective computing and gaming will be exploited to clearly understand users' experience, behaviour, and levels of immersion by leveraging the information captured by physiological sensors.

In future works, these AI applications will be further introduced to provide a dynamic mechanism that can change the HUD in real-time depending on the single-player responses.

\begin{acronym}
\acro{ai}[AI]{Artificial Intelligence}
\acro{dda}[DDA]{Dynamic Difficulty Adjustment}
\acro{ecg}[ECG]{electrocardiogram}
\acro{eda}[EDA]{electrodermal activity}
\acro{eeg}[EEG]{electroencephalogram}
\acro{eog}[EOG]{electrooculogram}
\acro{emg}[EMG]{electromyogram}
\acro{gsr}[GSR]{Galvanic Skin Response}
\acro{hud}[HUD]{Head's-Up Display}
\acro{ieq}[IEQ]{Immersive Experience Questionnaire}
\acro{ppg}[PPG]{photoplethysmogram}
\acro{tps}[TPS]{third person shooter}
\acro{ui}[UI]{User Interface}
\acro{vr}[VR]{Virtual Reality}
\end{acronym}

\bibliography{mybib}

\begin{thebibliography}{24}
\expandafter\ifx\csname natexlab\endcsname\relax\def\natexlab#1{#1}\fi
\providecommand{\url}[1]{\texttt{#1}}
\providecommand{\href}[2]{#2}
\providecommand{\path}[1]{#1}
\providecommand{\DOIprefix}{doi:}
\providecommand{\ArXivprefix}{arXiv:}
\providecommand{\URLprefix}{URL: }
\providecommand{\Pubmedprefix}{pmid:}
\providecommand{\doi}[1]{\href{http://dx.doi.org/#1}{\path{#1}}}
\providecommand{\Pubmed}[1]{\href{pmid:#1}{\path{#1}}}
\providecommand{\bibinfo}[2]{#2}
\ifx\xfnm\relax \def\xfnm[#1]{\unskip,\space#1}\fi
\bibitem[{E~Brown(2004)}]{Brown2004}
\bibinfo{author}{P.~C. E~Brown},
\newblock \bibinfo{title}{A grounded investigation of game immersion},
\newblock \bibinfo{journal}{Extended Abstracts of Human Factors in Computing
  Systems}  (\bibinfo{year}{2004}).
\bibitem[{Csikszentmihalyi(1975)}]{csikszentmihalyi1975beyond}
\bibinfo{author}{M.~Csikszentmihalyi},
\newblock \bibinfo{title}{Beyond boredom and anxiety. san francisco},
\newblock \bibinfo{journal}{CA, US: Jossey-Bass}  (\bibinfo{year}{1975}).
\bibitem[{Csikszentmihalyi(1990)}]{Csikszentmihalyi1990}
\bibinfo{author}{M.~Csikszentmihalyi}, \bibinfo{title}{Flow: The Psychology of
  Optimal Experience}, \bibinfo{year}{1990}.
\bibitem[{C~Jennet(2008)}]{Jennett2008}
\bibinfo{author}{.~A.~W. C~Jennet, A L~Cox},
\newblock \bibinfo{title}{Measuring and defining the experience of immersion in
  games},
\newblock \bibinfo{journal}{International Journal of Human-Computer Studies}
  \bibinfo{volume}{66} (\bibinfo{year}{2008}) \bibinfo{pages}{641--661}.
\bibitem[{M~Taylor(2017)}]{Taylor2017}
\bibinfo{author}{A.~C. M~Taylor},
\newblock \bibinfo{title}{Augmenting the hud: A mixed methods analysis on the
  impact of extending the game ui beyond the screen}  (\bibinfo{year}{2017}).
\bibitem[{Sanders~T(2010)}]{sanders&cairns}
\bibinfo{author}{C.~P. Sanders~T},
\newblock \bibinfo{title}{Time perception, immersion and music in videogames},
\newblock \bibinfo{journal}{Proceedings of HCI 2010 (HCI)}
  (\bibinfo{year}{2010}).
\bibitem[{Christou(2014)}]{christou2014interplay}
\bibinfo{author}{G.~Christou},
\newblock \bibinfo{title}{The interplay between immersion and appeal in video
  games},
\newblock \bibinfo{journal}{Computers in human behavior} \bibinfo{volume}{32}
  (\bibinfo{year}{2014}) \bibinfo{pages}{92--100}.
\bibitem[{Iacovides et~al.(2015)Iacovides, Cox, Kennedy, Cairns, and
  Jennett}]{Iacovides2015}
\bibinfo{author}{I.~Iacovides}, \bibinfo{author}{A.~Cox},
  \bibinfo{author}{R.~Kennedy}, \bibinfo{author}{P.~Cairns},
  \bibinfo{author}{C.~Jennett},
\newblock \bibinfo{title}{Removing the hud: the impact of non-diegetic game
  elements and expertise on player involvement},
\newblock in: \bibinfo{booktitle}{Proceedings of the 2015 Annual Symposium on
  Computer-Human Interaction in Play}, \bibinfo{year}{2015}, pp.
  \bibinfo{pages}{13--22}.
\bibitem[{Lages(2021)}]{lages2021opportunities}
\bibinfo{author}{W.~S. Lages},
\newblock \bibinfo{title}{Opportunities and challenges in immersive
  entertainment},
\newblock \bibinfo{journal}{Anais Estendidos do XX Simp{\'o}sio Brasileiro de
  Jogos e Entretenimento Digital}  (\bibinfo{year}{2021})
  \bibinfo{pages}{1037--1040}.
\bibitem[{Hughes and Jorda(2021)}]{hughes2021applications}
\bibinfo{author}{A.~Hughes}, \bibinfo{author}{S.~Jorda},
\newblock \bibinfo{title}{Applications of biological and physiological signals
  in commercial video gaming and game research: a review},
\newblock \bibinfo{journal}{Frontiers in Computer Science} \bibinfo{volume}{3}
  (\bibinfo{year}{2021}) \bibinfo{pages}{557608}.
\bibitem[{Bian et~al.(2016)Bian, Yang, Gao, Li, Zhou, Li, Sun, and
  Meng}]{bian2016framework}
\bibinfo{author}{Y.~Bian}, \bibinfo{author}{C.~Yang}, \bibinfo{author}{F.~Gao},
  \bibinfo{author}{H.~Li}, \bibinfo{author}{S.~Zhou}, \bibinfo{author}{H.~Li},
  \bibinfo{author}{X.~Sun}, \bibinfo{author}{X.~Meng},
\newblock \bibinfo{title}{A framework for physiological indicators of flow in
  vr games: construction and preliminary evaluation},
\newblock \bibinfo{journal}{Personal and Ubiquitous Computing}
  \bibinfo{volume}{20} (\bibinfo{year}{2016}) \bibinfo{pages}{821--832}.
\bibitem[{Tozman et~al.(2015)Tozman, Magdas, MacDougall, and
  Vollmeyer}]{tozman2015understanding}
\bibinfo{author}{T.~Tozman}, \bibinfo{author}{E.~S. Magdas},
  \bibinfo{author}{H.~G. MacDougall}, \bibinfo{author}{R.~Vollmeyer},
\newblock \bibinfo{title}{Understanding the psychophysiology of flow: A driving
  simulator experiment to investigate the relationship between flow and heart
  rate variability},
\newblock \bibinfo{journal}{Computers in Human Behavior} \bibinfo{volume}{52}
  (\bibinfo{year}{2015}) \bibinfo{pages}{408--418}.
\bibitem[{Setiono et~al.(2021)Setiono, Saputra, Putra, Moniaga, and
  Chowanda}]{setiono2021enhancing}
\bibinfo{author}{D.~Setiono}, \bibinfo{author}{D.~Saputra},
  \bibinfo{author}{K.~Putra}, \bibinfo{author}{J.~V. Moniaga},
  \bibinfo{author}{A.~Chowanda},
\newblock \bibinfo{title}{Enhancing player experience in game with affective
  computing},
\newblock \bibinfo{journal}{Procedia Computer Science} \bibinfo{volume}{179}
  (\bibinfo{year}{2021}) \bibinfo{pages}{781--788}.
\bibitem[{Picard(1999)}]{picard1999affective}
\bibinfo{author}{R.~W. Picard},
\newblock \bibinfo{title}{Affective computing for hci.},
\newblock in: \bibinfo{booktitle}{HCI (1)}, \bibinfo{organization}{Citeseer},
  \bibinfo{year}{1999}, pp. \bibinfo{pages}{829--833}.
\bibitem[{Mar{\'\i}n-Morales et~al.(2020)Mar{\'\i}n-Morales, Llinares,
  Guixeres, and Alca{\~n}iz}]{marin2020emotion}
\bibinfo{author}{J.~Mar{\'\i}n-Morales}, \bibinfo{author}{C.~Llinares},
  \bibinfo{author}{J.~Guixeres}, \bibinfo{author}{M.~Alca{\~n}iz},
\newblock \bibinfo{title}{Emotion recognition in immersive virtual reality:
  From statistics to affective computing},
\newblock \bibinfo{journal}{Sensors} \bibinfo{volume}{20}
  (\bibinfo{year}{2020}) \bibinfo{pages}{5163}.
\bibitem[{Tao et~al.(2006)Tao, Tan, and Picard}]{Tao2006}
\bibinfo{author}{J.~Tao}, \bibinfo{author}{T.~Tan},
  \bibinfo{author}{R.~Picard}, \bibinfo{title}{Affective computing and
  intelligent interaction}, volume \bibinfo{volume}{3784},
  \bibinfo{publisher}{Springer}, \bibinfo{year}{2006}.
\bibitem[{Kotsia et~al.(2013)Kotsia, Zafeiriou, and
  Fotopoulos}]{kotsia2013affective}
\bibinfo{author}{I.~Kotsia}, \bibinfo{author}{S.~Zafeiriou},
  \bibinfo{author}{S.~Fotopoulos},
\newblock \bibinfo{title}{Affective gaming: A comprehensive survey},
\newblock in: \bibinfo{booktitle}{Proceedings of the IEEE conference on
  computer vision and pattern recognition workshops}, \bibinfo{year}{2013}, pp.
  \bibinfo{pages}{663--670}.
\bibitem[{Moon et~al.(2022)Moon, Choi, Park, Choi, Hong, and Kim}]{Moon2022}
\bibinfo{author}{J.~Moon}, \bibinfo{author}{Y.~Choi},
  \bibinfo{author}{T.~Park}, \bibinfo{author}{J.~Choi}, \bibinfo{author}{J.-H.
  Hong}, \bibinfo{author}{K.-J. Kim},
\newblock \bibinfo{title}{Diversifying dynamic difficulty adjustment agent by
  integrating player state models into monte-carlo tree search},
\newblock \bibinfo{journal}{Expert Systems with Applications}
  \bibinfo{volume}{205} (\bibinfo{year}{2022}) \bibinfo{pages}{117677}.
\bibitem[{NielsenNorman-Group(2020)}]{nielsen}
\bibinfo{author}{NielsenNorman-Group},
\newblock \bibinfo{title}{10 usability heuristics for user interface design}
  (\bibinfo{year}{2020}).
\bibitem[{Federoff(2002)}]{Federoff2022}
\bibinfo{author}{M.~A. Federoff},
\newblock \bibinfo{title}{Heuristics and usability guidelines for the creation
  and evaluation of fun in video games}  (\bibinfo{year}{2002}).
\bibitem[{Wang et~al.(2016)Wang, Cheng, and Feris}]{Wang2016}
\bibinfo{author}{J.~Wang}, \bibinfo{author}{Y.~Cheng}, \bibinfo{author}{R.~S.
  Feris},
\newblock \bibinfo{title}{Walk and learn: Facial attribute representation
  learning from egocentric video and contextual data},
\newblock in: \bibinfo{booktitle}{Proceedings of the IEEE Conference on
  Computer Vision and Pattern Recognition (CVPR)}, \bibinfo{year}{2016}.
\bibitem[{Gannouni et~al.(2021)Gannouni, Aledaily, Belwafi, and
  Aboalsamh}]{gannouni2021emotion}
\bibinfo{author}{S.~Gannouni}, \bibinfo{author}{A.~Aledaily},
  \bibinfo{author}{K.~Belwafi}, \bibinfo{author}{H.~Aboalsamh},
\newblock \bibinfo{title}{Emotion detection using electroencephalography
  signals and a zero-time windowing-based epoch estimation and relevant
  electrode identification},
\newblock \bibinfo{journal}{Scientific Reports} \bibinfo{volume}{11}
  (\bibinfo{year}{2021}) \bibinfo{pages}{7071}.
\bibitem[{Redmond(2013)}]{redmond2013trans}
\bibinfo{author}{C.~Redmond},
\newblock \bibinfo{title}{Trans-thoracic impedance measurements in patient
  monitoring},
\newblock \bibinfo{journal}{EDN Network}  (\bibinfo{year}{2013}).
\bibitem[{Sch{\"o}ber and Stadtmann(2020)}]{schober2020fortnite}
\bibinfo{author}{T.~Sch{\"o}ber}, \bibinfo{author}{G.~Stadtmann},
\newblock \bibinfo{title}{Fortnite: The business model pattern behind the
  scene},
\newblock \bibinfo{journal}{European University Viadrina Frankfurt (Oder)
  Department of Business Administration and Economics Discussion Paper}
  (\bibinfo{year}{2020}).

\end{thebibliography}

\end{document}